\documentclass[aps,prb, twocolumn,groupedaddress]{revtex4}

\usepackage{graphicx}
\usepackage{amssymb}

\newcommand{\ybinag}{YbIn${_{1-x}}$Ag${_{x}}$Cu${_{4} }$}
\newcommand{\ybth}{YbIn${_{.7}}$Ag${_{.3}}$Cu${_{4} }$}
\newcommand{\ybin}{YbInCu${_{4} }$}
\newcommand{\ybag}{YbAgCu${_{4} }$}
\newcommand{\tv}{${T_{v} }$}

\newcommand{\tk}{${T_{K} }$}
\newcommand{\wn}{${cm^{-1} }$}
\newcommand{\sigone}{$\sigma_{1}(\omega)$}
\newcommand{\EF}{$E_{F}$}

\newcommand{\eV}{$eV$}

\newcommand{\x}{$x$}

\newcommand{\K}{$K$}

\newcommand{\twiddle}{$\sim$}
\newcommand{\f}{${f}$}

\newcommand{\s}{\hspace{.05cm}}
\newcommand{\etal}{\textit{et al}}

\begin{document}

\title{The Kondo Dynamics of \ybinag}

\author{Jason N. \surname{Hancock}}
\email{jason@physics.ucsc.edu}

\author{Tim \surname{McKnew}}
\author{Zack \surname{Schlesinger}}
\affiliation{Physics Department, University of California Santa Cruz, Santa Cruz, CA 95064, USA}

\author{John L. \surname{Sarrao}}
\affiliation{Los Alamos National Laboratory, Mail Stop K764, Los Alamos, New Mexico 87545, USA}

\author{Zach \surname{Fisk}}
\affiliation{National High Magnetic Field Laboratory, Tallahassee, Florida 32310, USA}
\affiliation{Department of Physics, University of California Davis, Davis CA  95616, USA}

\date{\today}
\begin{abstract}
We present  an infrared/optical study of the dynamics of the strongly correlated electron system \ybinag\ as a function of doping and temperature for  \x\ ranging from 0 to 1, and T between 20 and 300 K.  This study
reveals information about the unusual phase transition as well as the phases themselves. Scaling relations emerge from the data and are investigated in detail using a periodic Anderson model based calculation. We also provide a picture in which to view both the low and high-energy  \x\ -dependent features of the infrared data, including identification of high energy, temperature dependent features.
\end{abstract}
\maketitle

When exploring the complex terrain of correlated electron phases, one is often interested in the phase boundaries separating systems which are controlled by different physics. In the low doping regime (\x$<$0.2) of \ybinag, a temperature-driven, first-order electronic phase transition separates two phases which are characterized by widely disparate effective energy scales (\tk) associated with the effects of hybridization between the localized Yb \f-electrons and the itinerant conduction electrons. This rapid change provides both an opportunity to identify and isolate features associated
with many-body hybridization physics and raises fundamental questions regarding the origin of the transition itself.

The valence transition in \ybin\ has been examined with a variety of experimental techniques 
\cite{lawrence2,sarrao2,cornelius97,immer,lawrence3,kindler,teresa}, and exhibits some similarities\cite{nowik1,nowik2,sarrao4} to phase transitions observed in certain rare-earth intermetallic compounds\cite{vandereb,allen1}.
Samples of high quality can be made, and provide the only known example of a valence transition occurring at ambient pressure in a stoichiometric sample. The synthesis of high quality single crystal samples has allowed the detailed study of the phase diagram of \ybinag, and continued effort has revealed a tendency of the system to order ferromagnetically as the phase transition temperature is driven to zero either by pressure\cite{mitsuda02,park04} or Y doping\cite{mitsuda02,mito03}. The presence of interesting phase competition in a system where the sample chemistry is well controlled enhances our interest in this correlated electron system.

\begin{figure}[htbp]
\centerline{\scalebox{.3}{\includegraphics{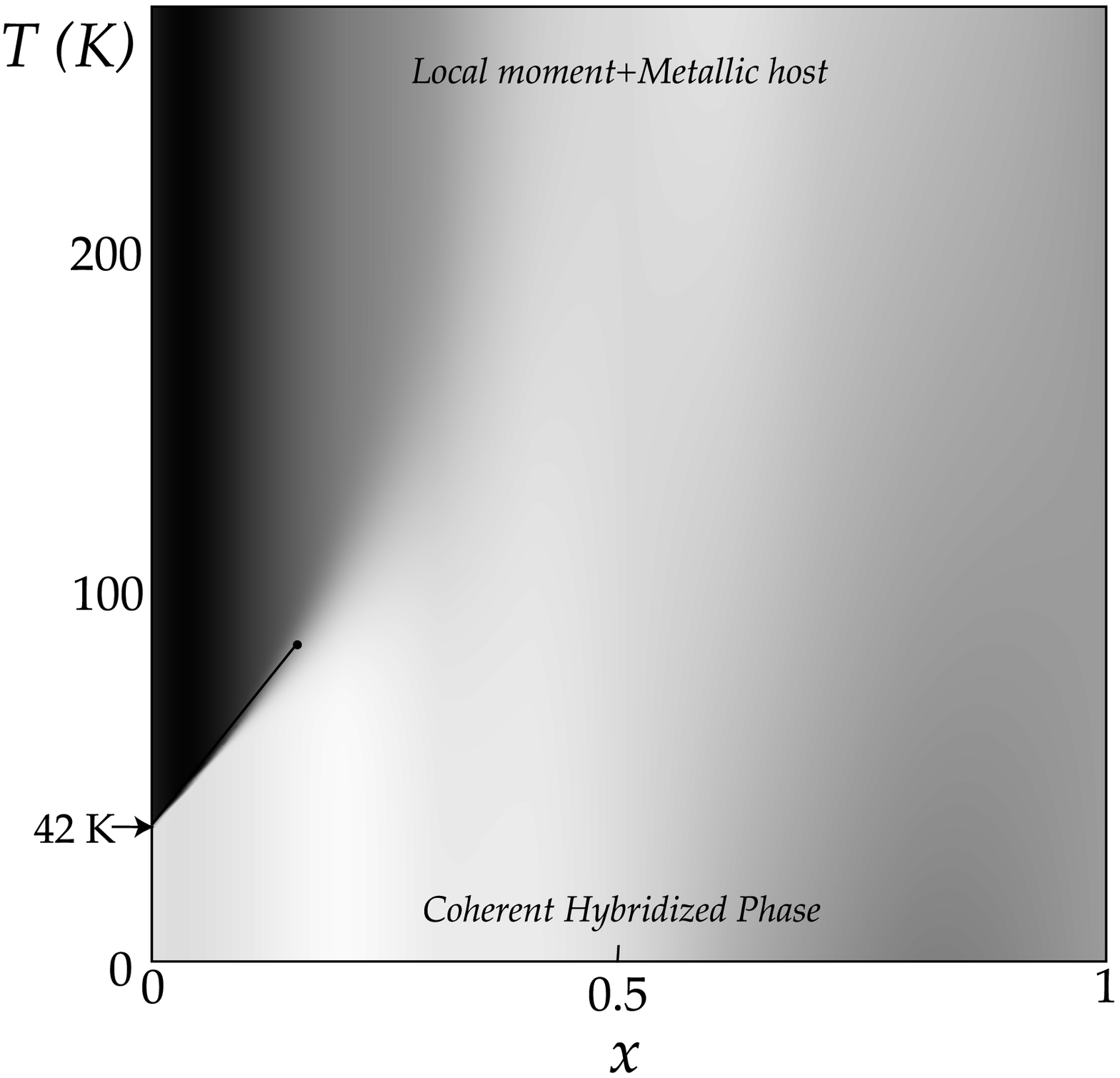}}}
\caption{Schematic phase diagram of \ybinag. Lighter shading indicates a larger \tk.}
\label{fig:pd}
\end{figure}

Figure \ref{fig:pd} shows a schematic representation of the phase diagram of \ybinag. At low doping, a line of first order phase transitions separates a low temperature, mixed-valent phase and a high temperature local moment phase. In \ybin\ (\x=0) at high temperature ($T>T_v=42\s K$), the magnetic response exhibits a Curie-Weiss form with magnitude appropriate to $j=7/2$ moment of Yb and a small Weiss temperature\cite{svechkarev} $\Theta_W\sim-13\s K$, which constrains the effective Kondo temperature appropriate in this high temperature range to be comparably small in magnitude. The resistivity in this temperature range\cite{sarrao2} is very high for a metal ($\rho_{\textrm{dc}}\simeq150\mu\Omega$-$cm$) and together with Hall measurements indicate a very low (hole) carrier concentration\cite{sarrao2}. Lowering the temperature through the first order phase transition has dramatic effects on both the spin and charge response. For $T<T_v$, the magnetic susceptibility drops considerably into a temperature independent (Pauli paramagnetic) form, with a magnitude indicating a dramatic increase in the Kondo temperature. In concert the carrier concentration increases markedly and the resistance drops\cite{sarrao2}. Extensive experimentation has revealed substantial changes in specific heat\cite{sarrao3, nowik1, hauser2}, elastic constants\cite{sarrao3, kindler}, single-particle (photoemission)\cite{dallera,joyce2,weibel, lawrence1,reinert}, two-particle (optical)\cite{garner00, hancock04, antonov,marab3} and neutron\cite{lawrence3} spectroscopies.

Changes associated with the phase transition in \ybin\ are generally interpreted as an effective screening of the localized Yb 4\f\ moments in the low temperature phase. This is associated with an order of magnitude increase in Kondo coupling scale (\tk\twiddle17\s\K\ for $T>T_v$; \tk\twiddle300\s\K\ for $T<T_v$). Universal scaling observed\cite{immer,sarrao5} in the $B$-$T$ plane was shown\cite{dzero1,mushnikov03} to be consistent with the scenario of a transition-induced screening of the Yb moment.

Doping Ag on the In site serves to stabilize the low temperature phase, driving the phase transition temperature upward\cite{sarrao2}. Experimentally, the transition becomes less sharp with doping and it is estimated\cite{sarrao2} that the transition disappears completely at the critical concentration $x_c\simeq0.2$. Concentration values $x>x_c$ display behavior more typical of heavy fermion systems, where coherent heavy metallic quasiparticles responsible for the low energy physics have an enhanced mass as high as $m^*\simeq50m_e$.

Much of the phenomenology of \ybinag\ is reflective of periodic Anderson model (PAM) physics, however, the phase transitions known to exist\cite{schroder} in the context of the PAM are associated with a competition between magnetic order and Fermi liquid phases and occur in a region of weak coupling. This region is far from the mixed-valent fixed point, where the coupling is strong or intermediate, and these transitions therefore seem unlikely to be directly relevant to \ybinag.  In the region of intermediate to strong coupling the PAM exhibits gradual crossover to a coherent state as temperature is reduced, and does not, to our knowledge,  include phase transitions. One may therefore ask whether the line of phase transitions occurring at low \x\ in \ybinag, which are electronic and not structural in origin, imply the need for an additional interaction term in the minimal model Hamiltonian beyond those normally included in the PAM. An intriguing possibility is that electron-electron interactions, which can be very effective in bringing about transitions in Mott-Hubbard systems, may play a role in \ybinag. Investigations along these lines have been pursued by Giamarchi et al.\cite{giamarchi93} and Zlatic and Freericks\cite{zlatic01,freericks03,fkrmp}.

In this paper, we present infrared-frequency dynamics derived from measurements of the reflectivity of single crystal \ybinag, with Ag compositions \x=0, 0.3, 0.5, 0.75, and 1.0. This systematic study allows us to identify and explore scaling behavior of the infrared signature of the Kondo resonance. In the \x=0 system (\ybin), these measurements\cite{garner00} revealed a mid-infrared excitation prominent primarily in the low temperature phase\cite{garner00,hancock04} ($T<T_v=42\s K$). This may be understood as a consequence of the presence of hybridized quasiparticle bands\cite{garner00,hancock04,dordevic,degiorgi01}. Here we explore in detail the phenomenology of this excitation. Within the low temperature phase we use modeling to investigate the scaling properties of this infrared excitation. In addition, we present interpretations of the higher frequency features based on 
systematic trends of spectral features with Ag substitution (hole doping) and propose an energy level scheme in which to view the high-frequency spectroscopic results.

Our measurements cover the frequency range from ${40\s cm^{-1}}$ to ${50,000\s cm^{-1}}$ with detailed temperature dependent data taken between ${40\s cm^{-1}}$ and ${23,000\s cm^{-1}}$. Each sample was mounted into a recess custom-machined into a brass disk. Each disk was polished (minimally) such that a smooth, flat sample surface sits parallel to a similarly flat, polished portion of the brass mounting disk. Ag film was then evaporated on the bare brass surface, creating a reference mirror for use in the temperature dependent reflectivity measurements. Soon after evaporation, the sample disk was mounted inside a continuous-flow He$^4$ cryostat with a custom switching mechanism designed to expose either Ag or sample surfaces to the beam. To insure that the reference mirror and sample surface are coplanar and flat, a laser alignment procedure was used on the mounted sample disk. Frequency dependent spectra were then taken using a combination of Fourier transform and grating spectrometers. At each temperature and composition, spectra were taken for Ag and sample exposed through the cryostat window and the the ratio at each frequency used to determine the reflectivity $R(\omega)$.

In addition to the temperature dependence, room temperature reflectivity spectra were also taken in the range $12,500\s cm^{-1}<\omega<50,000\s cm ^{-1}$ with an optical setup which measures absolute refelctivity without the use of a reference mirror. The spectra taken by this method and those spectra taken using the cryostat (and with Ag reference) generally agree well in the region of overlap ($12,500\s cm^{-1}<\omega<23,000\s cm ^{-1}$). Through this comparison, we estimate the uncertainty of the absolute value of the reflectivity in the temperature dependent measurements to be less than 1\%.

A Kramers-Kronig transform is applied to the measured reflectivity in order to determine the frequency dependent reflection phase shift\cite{dressel, wooten}. The magnitude and phase of the reflectivity are then used in order to determine the dynamical conductivity \sigone\ (and the dielectric function $\epsilon_1(\omega)$).

For the purposes of the transform, Hagen-Rubens terminations ($1-R(\omega)\propto\sqrt{\omega}$) are used below ${40 \s cm^{-1}}$. At high frequency (above ${50,000\s cm^{-1}}$) each reflectivity spectrum is extrapolated to a common value of 0.08 at ${120,000\s cm^{-1}}$ and then continued as a constant to ${200,000\s cm^{-1}}$. Between ${200,000\s cm^{-1}}$ and ${400,000\s cm^{-1}}$, an ${\omega^{-2}}$ form is used for ${R(\omega)}$ to represent the non-constant reflectivity expected as a result of deep core level excitations\cite{wooten}. At still higher frequnecies, the free-electron form ${\omega^{-4}}$ is assumed.

We have experimented with a number of termination protocols including other common values and coalescence frequencies as well as constant extrapolations above ${50,000\s cm^{-1}}$. These show convincingly that our results regarding trends in the \x\ and \tk\ dependence of \sigone\ below ${10,000\s cm^{-1}}$ are not significantly influenced by any of the extrapolations above ${50,000\s cm^{-1}}$. The conductivity above 10,000\s\wn\ but below 20,000\s\wn\ is influenced by the detailed extrapolation by about 5\% for reasonable extrapolation protocols.

In \ybin, where a first-order phase transition occurs at finite temperature, we have been careful to control the potential hysteresis effects\cite{sarrao2,kindler} usually associated with first order phase transitions by keeping the thermal cycling through the phase transition to a minimum while collecting the optical data. We found that hysteresis effects are observable in the infrared spectrum, and are especially pronounced in the frequency range $1000<\omega<7000\s cm^{-1}$ (see Figure \ref{fig:x0}). For this reason, our data for several ranges were retaken on fresh samples.

\section{Results}

\begin{figure}[htbp]
\centerline{\scalebox{.4}{\includegraphics{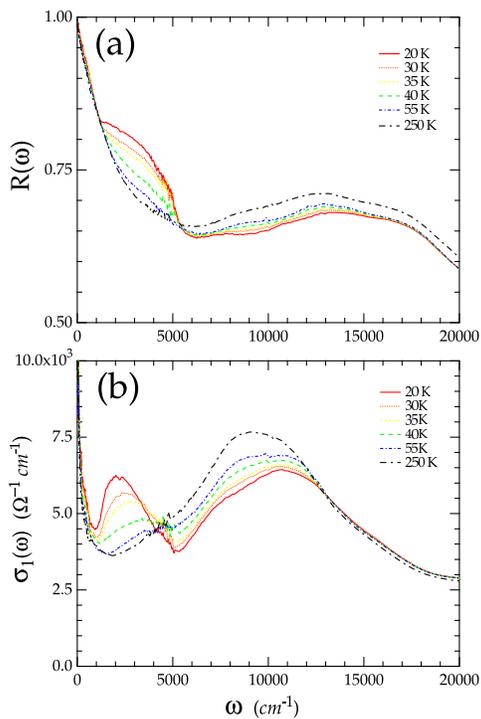}}}
\caption{(Color online) The real part of the optical conductivity of \ybin\ at various temperatures.}
\label{fig:x0}
\end{figure}

Figure \ref{fig:x0} shows the frequency-dependent reflectivity and infrared conductivity \sigone\ of \ybin\ at temperatures below and above the $T_v\simeq42K$ phase transition temperature. At high temperature (250\s\K), \sigone\ consists of a narrow free-carrier (Drude-like) contribution clearly seen in the far-infrared spectrum, due to the presence of mobile carriers. This interpretation is consistent with Hall\cite{figueroa98} and resistivity measurements\cite{sarrao2}. $\omega\simeq6000cm^{-1}$ marks a clear onset of a set of strong interband transitions extending upward into the visible spectral range. 

At the lowest temperature (20\s\K), a significant decrease in conductivity in the interband region ($\omega>4000\s cm^{-1}$) accompanies a substantial increase in the conductivity below 4000\s\wn\ in the form of a well-defined peak centered around 2000\s\wn. The development of this peak is highly correlated with the phase transition (\tv$=$42\s\K) and is a prominent and essential feature of the low-$T$ phase of \ybin. The connection between this peak and the physics of hybridization is an important theme of the present work.

\begin{figure}[htbp]
\centerline{\scalebox{.4}{\includegraphics{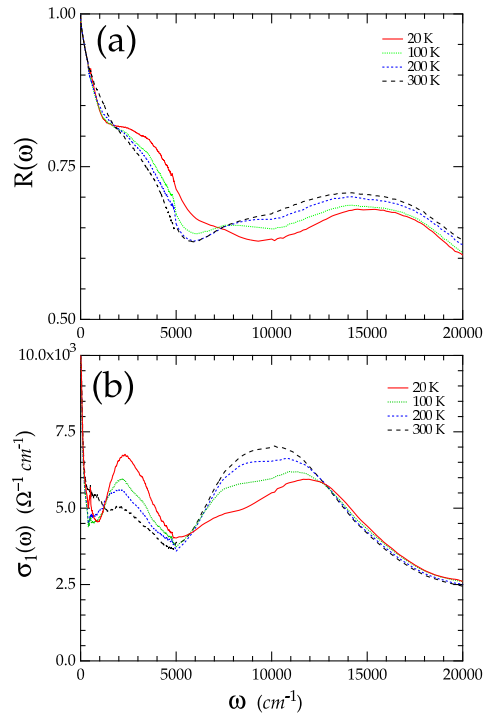}}}
\caption{(Color online) The real part of the optical conductivity of \ybth\ at various temperatures.}
\label{fig:x3}
\end{figure}

\begin{figure}[htbp]
\centerline{\scalebox{.4}{\includegraphics{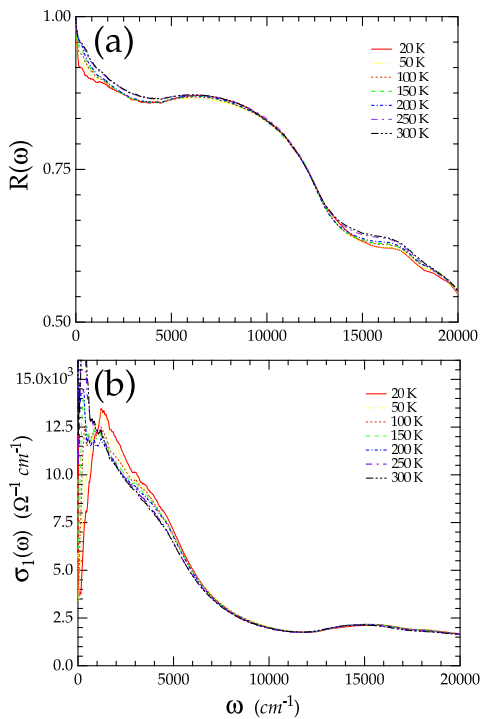}}}
\caption{(Color online) The real part of the optical conductivity of \ybag\ at various temperatures.}
\label{fig:x1}
\end{figure}

Figure \ref{fig:x3} shows optical data for the \x=0.3 system. Trends similar to those in low temperature \ybin\ are apparent, with a depletion of weight around 8000\s\wn, accommodated by a replenishing at lower frequencies. Vestiges of the 2000\wn\ peak persist at temperatures as high as 300\s\K, an energy scale much higher than the extrapolated value of the phase transition temperature for this composition \tv(\twiddle100\K), but still lower than the Kondo scale appropriate to the low temperature phase \tk\twiddle360\K.

Figure \ref{fig:x1} shows the similar plots for \ybag. The behavior displayed by the conductivity shows behavior more typical of a heavy fermion system, with moderate temperature dependence and spectral weight which is approximately conserved below 0.5\s\eV. There is a continuous evolution of the 2000\s\wn\ feature, activated by crossing the valence transition at low \x, and the more familiar phenomenology found in the low frequency response of the heavy fermi system \ybag.

\begin{figure}
\begin{center}
\includegraphics[width=3in]{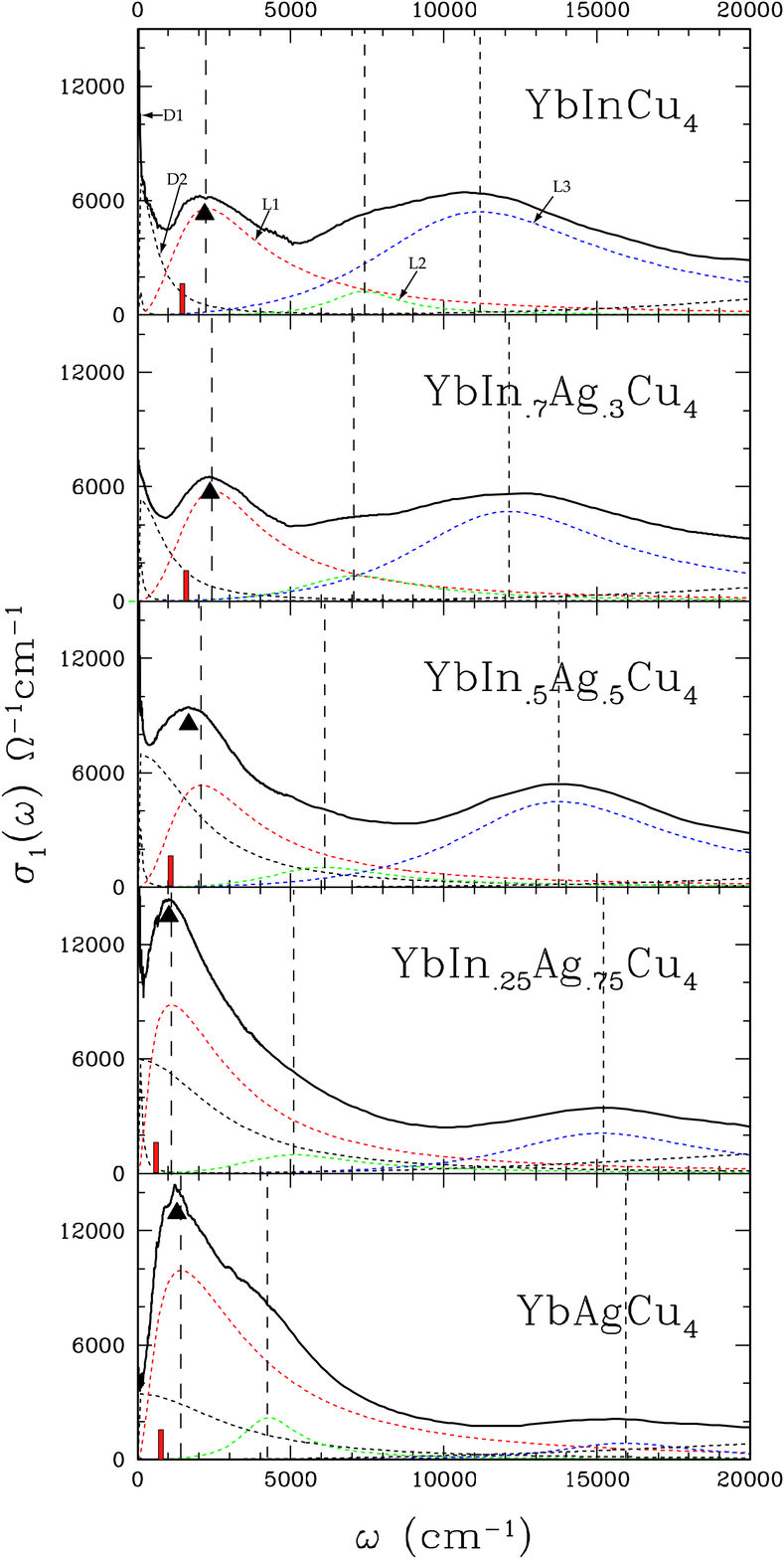}
\caption{(Color online) Infrared conductivity \sigone\ versus $\omega$ at 20\s\K\ for the five \x\ values studied. Also shown are the components of a Lorentzian-Drude fit as described in the text. The dashed vertical lines are a guide to the eye and correspond to the center frequencies of the fit components when \x=0.}
\label{fig:allx}
\end{center}
\end{figure}

Figure \ref{fig:allx} profiles the \x\ dependence of \sigone\ at low temperatures. The 2,000\s\wn\ peak in \ybin\ undergoes a complex shifting behavior as \x\ is increased, blueshifting slightly when \x=0.3, then redshifting upon further doping, reaching a minimum peak frequency when \x=0.75, before blueshifting again as \x\ continues to 1. The strength of this feature is also influenced by \x\ in a nontrivial way, discussed further below.

At higher frequency, the large hump feature centered on 11,000\s\wn\ in \ybin\ monotonically blueshifts and decreases in overall strength as \x\ is increased. Further, inflection points around 6000\s\wn\ and 9000\s\wn\ redshift slightly upon doping to the \x=0.3 and \x=0.5 systems. These inflection points are not discernable for \x=0.75, but reappear at low frequency (3000\s\wn\ and 5500\s\wn) in the \x=1 system.

The \x-dependent profiling of the low temperature conductivity is an important part of our experimental results, allowing clear identification of systematic changes of the low temperature electrodynamics as a function of an external control parameter. The discussion begins with identification of systematic trends in the 2000\s\wn\ feature followed by a discussion of interband features in a later section.

\section{Results, \x\ dependence}
\label{sec:lo}

\begin{figure} 
\centerline{\scalebox{.5}{\includegraphics{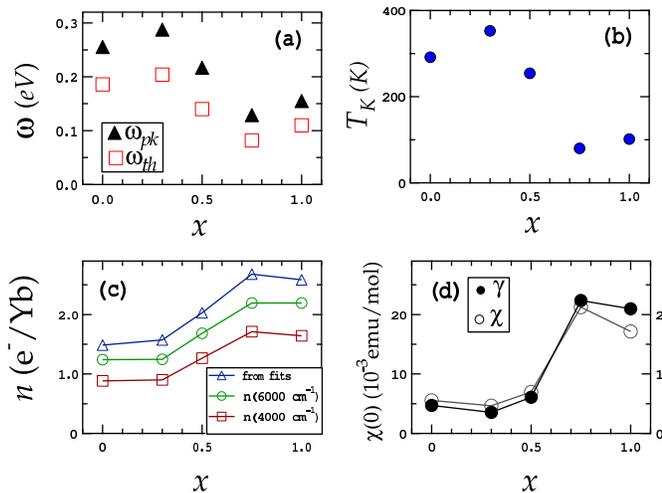}}}
\caption{(Color online) (a) The characteristic frequencies, ${\omega_{pk}}$ (triangles) and ${\omega_{th}}$ (boxes), and (b) the Kondo temperature as a function of \x. (c) The spectral weights, ${n(4000\s cm^{-1})}$ (boxes) and ${n(6000\s cm^{-1})}$ (circles) and the results of a combined Drude/Lorentz fitting (triangles) of the low frequency conductivity. (d) The low temperature susceptibility (open circles) and Sommerfeld coefficient (solid circles) as a function of \x. \tk\ in (b) is related to $\chi(0)$ in (d). Spectral weights are computed assuming a band mass of 4$m_e$. }\label{fig:ntw}
\end{figure}

In this section, we focus on the \x\ dependent changes of the 2000\s\wn\ feature in order to explore the significance of this behavior to the correlated electron physics of hybridization and the periodic Anderson model. First, we quantify the \x\ dependent trends of Figure \ref{fig:allx} and draw important conclusions through the comparison to previously published thermodynamic data. The inferred relationships are then explored below in sections \ref{sec:kg} and \ref{sec:pam} where we consider an interpretation based on the periodic Anderson model.

Figure \ref{fig:ntw}a shows the frequency of the 2000\s\wn\ feature versus \x\ as determined in two ways. The black triangles mark the frequency of the peak in \sigone\ (also marked in Figure \ref{fig:allx}). Alternatively, a threshold frequency can be extracted from a fit of the conductivity to a calculation based on the low energy dispersion of the periodic Anderson model (PAM), discussed below in Section \ref{sec:kg}.

In addition to examining the frequency of the peak as a function of \x, we can also look at the strength of the 2000\s\wn\ feature. We quantify this characteristic through the spectral weight, defined as the integrated intensity of \sigone\ over a low frequency interval:
\begin{equation} n(\omega)=\frac{2 m}{\pi
e^{2}}\int_{0^{+}}^{\omega}\sigma_1(\omega^{\prime})\mathrm{d}\omega^{\prime}
\label{eqn3:sw} \end{equation}
where $m$ represents a bare band mass. The lower limit is chosen to be nonzero ($0^+=50\s cm ^{-1}$) in order to exclude from the strength estimate the comparatively minute contribution of the free carrier (Drude) response. The upper limit of integration is chosen to encompass the 2000\s\wn\ peak without including the \x-dependence of the high frequency interband contributions. Neither integration limit is critical; in fact a lower limit of 0 and upper limits between anywhere between 3000\s\wn and 8000\s\wn\ produce similar \x\ dependence. $n(4000\s cm^{-1})$ and $n(6000\s cm^{-1})$ are shown in Figure \ref{fig:ntw}c.

As an alternative to this simple integral calculation of the strength, we can fit the complex conductivity ($\sigma=\sigma_1+i\sigma_2$) with a sum of Lorentzian and Drude response functions\cite{dressel,wooten}:
\begin{equation}
\sigma(\omega)=\sum_j\frac{\omega_{P,j}^2}{4\pi}\frac{\omega}{i(\omega_{j}^2-\omega^2)+\omega\Gamma_j}.
\label{eq:lo}
\end{equation}
The constituents of the fits include: one narrow Drude (D1, $\Gamma\sim10-40\s cm^{-1}$) contribution to represent the free carrier peak, a wide Drude (D2, $\Gamma>800\s cm^{-1}$), and a Lorentz oscillator (L1) in the vicinity of the 2000\s\wn\ feature which, as we discuss below, relate to the Kondo resonance. There are in addition two Lorentz oscillators (L2 and L3, around 7,300\s\wn\ and 11,000\s\wn\ for \x=0) to represent the infrared interband conductivity, and two wide Lorentz oscillators at ultraviolet frequencies ($\omega>30,000\s cm^{-1}$) representing the conductivity in that range. These fit components are labeled in Figure \ref{fig:allx}a.

Previous work\cite{tahvildar} has explicitly demonstrated that the calculated lineshape of the optical signature of the Kondo resonance is intrinsically non-Lorentzian, and furthermore demonstrated the viability of fits which combine Drude and Lorentz terms to represent the Kondo resonance. The combination of the contributions D2 and L1 reasonably fit the conductivity in the range of the 2000\s\wn\ peak. The strength from that combination is shown by the triangles in Figure \ref{fig:ntw}c. This determination of the strength exhibits an \x\ dependence similar to the simpler integral representations of the strength. This makes one confident that the ${n}$ versus \x\ dependence shown here is an essential characteristic of the data, and independent of any of the detailed choices we have made in the analysis. $n$ determined by the methods discussed here is presented in Figure \ref{fig:ntw}c.

It is interesting to compare the \x-dependent trends inferred from previously published thermodynamic measurements with those from our optical data. Figure \ref{fig:ntw}b shows the Kondo temperature of low-$T$ phase \ybinag\ as deduced by Cornelius\cite{cornelius97} \etal\ from fitting the measured magnetic susceptibility to the numerically calculated result of the $j=\frac{7}{2}$ Coqblin-Schreiffer model\cite{cornelius97,rajan}. Figure \ref{fig:ntw}d shows that the trend in the \x\ dependence of $\chi(T=0)$\footnote{In the $N$-fold degenerate Anderson impurity model, the value of the low temperature susceptibility is related\cite{hewson} simply to the Kondo temperature \tk\ by $\chi(0)=(g\mu_B)^2j(j+1)w_N/3k_BT_K$, where $w_N$ is given by Equation \ref{eqn:wn}.} generally agrees with the corresponding trend in the Sommerfeld coefficient $\gamma$, implying a Wilson ratio within 10\%\ of the value ($\mathcal{R}=\frac{8}{7}$) expected for a $j=\frac{7}{2}$ Anderson impurity\cite{hewson,cornelius97}. Thus the \tk\ values inferred from the susceptibility analysis reasonably represent the effective energy scale relevant to the onset of strong-coupling Kondo physics within the low temperature phase.

The complicated \x\ dependence of \tk\ is not understood, and may be the result of an interplay of band structure, chemical pressure, screening, disorder, and other many-body effects. While resolving the detailed cause of this complex \x\ dependence presents a subject for future work, out focus here is the relationship between the \x\ dependence of \tk\ and that of the optical data; the similar form of the \x\ dependent electrodynamic (Figures \ref{fig:ntw}a and \ref{fig:ntw}c) and thermodynamic quantities (Figures \ref{fig:ntw}b and \ref{fig:ntw}d) indicates a common source. We show below that this is rooted in the strongly correlated electron physics of hybridization.

\section{Modeling, PAM Dispersion}
\label{sec:pam}
\begin{figure} \centerline{ \scalebox{.4}{\includegraphics{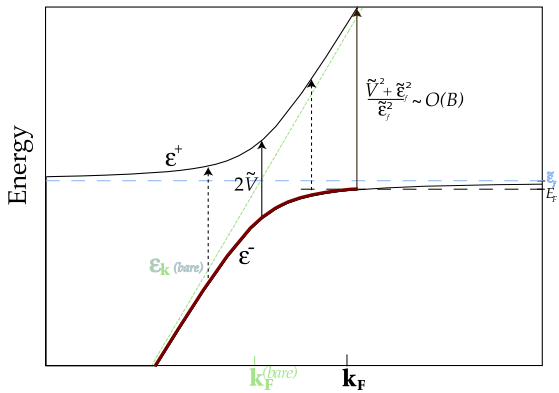}}}
\caption{(Color online) The PAM dispersion relations. Vertical arrows indicate possible optical transitions. The horizontal dashed lines represent \EF\ and $\tilde{\epsilon_f}$. The light diagonal line represents the unrenormalized dispersion of the conduction carriers.}
\label{fig:qpd}
\end{figure}

We can make progress toward eliciting the relationships suggested in Figures \ref{fig:ntw} by examining an interpretation of the 2000\s\wn\ feature in the context of the periodic Anderson model (PAM). Complementary to the work of other authors\cite{tahvildar,jarrell95,vidhyadhiraja03,georges96,zlatic01,freericks03}, which focus the rigorous techniques of many-body theory directly toward the underlying Hamiltonian, we will use a simplified approach based on effective low energy (near-\EF) PAM dispersion relations\cite{edwards,hewson}. These dispersion relations ($\epsilon^+$ and $\epsilon^-$ in Figure \ref{fig:qpd}) provide the basis for a simple unifying picture in which much of the low energy phenomenology of heavy fermion materials  can be viewed, including the mass enhancement, aspects of magnetism, and transport measurements\cite{millis87a,tannous}. In Figure \ref{fig:qpd}, the vertical extent of the plot is of order 1\s\eV\ and the (singly occupied) \f-electron level is below the bottom plot boundary. The light dashed lines indicate the bare (unhybridized) conduction electron dispersion.

In a system with no hybridization, the conduction electrons are the dominant influence on the transport properties such as thermopower and resistivity, while at the same time provide a temperature-independent Pauli-paramagnetic contribution to the magnetic susceptibility. The \f-electrons, on the other hand, are localized and as a result contribute very little to the transport properties, but play a major role in magnetism, contributing a Curie $1/T$ term to the susceptibility. Inclusion of the hybridization and on-site Coulomb repulsion terms complicates this independent particle picture considerably and the new eigenstates become nontrivial admixtures of the states of pure \f\ and conduction electron character.

The PAM quasiparticle dispersion relations provide a venue through which to explore the effects of nonzero hybridization on the phase space of excitations. At energies far from the chemical potential, the upper and lower bands, ${\epsilon^{+}}$ and ${\epsilon^{-}}$, follow closely the unrenormalized free carrier dispersion. There the main effect of the interaction and hybridization is to provide a channel for relaxation of the conduction states, \textit{i.e.} a broadening of the spectral function along the dispersion curves. At lower energies, these relaxation effects are reduced by the phase-space-constraining presence of a filled Fermi sea, however, renormalization also opens the Fermi surface ($k_F^{(bare)}\rightarrow k_F$ and the bands flatten) to accommodate the \f-electron weight projected up to the Fermi level. This reorganization of the bands in the vicinity of the Fermi level is due to many-body interactions, the strength of which is characterized by the parameter ${\tilde{V}}$. The resultant narrow peak in the density of states, called the Kondo, or Abrikosov-Suhl, resonance\footnote{Strictly speaking, there appears a peak in the $DOS$ in the impurity case. The periodic system differs in that an additional \textit{indirect} gap within this peak is present\cite{georges96}.} is central to the understanding of heavy fermion and mixed-valent phenomenology\cite{degiorgi99,hewson, jarrell95}.

From the point of view of optical probes, a key effect of renormalization on the optical response is to create the possibility for vertical transitions from filled states below \EF, across a \textit{direct} gap, and into unoccupied levels above \EF, as illustrated by the vertical arrows of Figure \ref{fig:qpd}. The threshold for these transitions (short arrow) occurs at a frequency ${\omega=2\tilde{V}}$, where ${\tilde{V}}$ is the hybridization strength renormalized by the on-site \f-electron repulsion. This energy scales with the Kondo temperature as\cite{coleman, coxcomm,millis87a,millis87b,grewe84}
\begin{equation}\tilde{V}=\sqrt{T_{K} B}\label{eq:tkb}\end{equation}
where ${B}$ is a model parameter related to the conduction electron bandwidth.
At threshold, the nesting condition for the upper and lower bands is met ($\nabla_{\textbf{k}}\epsilon^+=\nabla_{\textbf{k}}\epsilon^-$), leading to a very high joint density of states for vertical quasiparticle transitions and hence a strong peak in the conductivity. %We will explore this consequence quantitatively below and compare the results to the observed trends of Figure \ref{fig:ntw}.

At frequencies larger than the threshold frequency there are two distinct
contributions to the conductivity: one originating from levels
inside the unrenormalized Fermi surface (${k_{<}}$ below); the other from the states
occupied as a result of renormalization, \textit{i.e.}, outside the unrenormalized
Fermi surface (${k_{>}}$ below). An example of two such transitions with the same frequency $\omega$ are indicated by dashed arrows in Figure \ref{fig:qpd}. Transitions involving both of these sets of quasiparticles are important and must be counted independently in the determination of the total optical conductivity, as discussed further below.

In addition to the threshold frequency, another, higher frequency scale appears which is relevant to the electrodynamic response. This higher frequency scale corresponds to the vertical transition (long arrow, Figure \ref{fig:qpd}) which occurs from states \textit{on} the Fermi surface (\textit{i.e.} the locus of points which divides the set of occupied and unoccupied \textbf{k} states). Vertical transitions involving higher \textbf{k} states cannot occur because both initial and final states are unoccupied when $k>k_F$, and hence one expects a drop in the conductivity at this frequency. For a linearly dispersing conduction band, direct calculation reveals that the frequency of the last allowed transition is equal to
\begin{equation}
\label{eqn:OFS}
\Omega_{FS}=\frac{\tilde{V}^{2}+\tilde{\epsilon_{f}}^{2}}{\tilde{\epsilon_{f}}}.
\end{equation}
The identification of $\tilde{\epsilon_{f}}$ with \tk\ (discussed further below), together with Equation \ref{eq:tkb} implies that this scale is of order the conduction electron bandwidth, $B$. For high energy transitions, band edge final states can be reached and the linear approximation to the conduction band is likely to become poor. Quasiparticle transitions in this frequency range may be influenced by the details of the underlying band structure.
%Despite these ambiguity associated with the precise form of this energy scale, we will refer later to equation \ref{eqn:OFS} as a convenience in an estimate of the relative strength of the optical transitions discussed here. In a further section we will amend these 

\section{Modeling, Kubo-Greenwood Analysis}
\label{sec:kg}

With these considerations of the phase space for optical transitions in mind, we can proceed with an analysis using the Kubo-Greenwood formula\cite{dressel}:
\begin{equation}
\sigma_{1}(\omega) = \frac{\pi e^{2}}{m^{2}\omega}
\sum_{\ell,\ell^{\prime}}JDOS_{\ell,\ell^{\prime}}(\omega)|\textbf{p}_{\ell,\ell^{\prime}}|^{2}
\label{eq:conduct1}
\end{equation}
where ${|\textbf{p}_{\ell,\ell^{\prime}}|}$ denotes the dipole matrix element
connecting electronic bands ${\ell}$ and ${\ell^{\prime}}$, and
${JDOS_{\ell,\ell^{\prime}}(\omega)}$ is the corresponding joint density of states. Applying this formula to hybridizing quasiparticles (as though they were electrons) allows an exploration of the phenomena of the mid-infrared conductivity in the context of the PAM. In that case, the two relevant bands are $\epsilon^{+}$ and $\epsilon^{-}$, which in this model approach leads to:
\begin{equation}
\sigma_{pam}(\omega) = \frac{e^{2}}{4 \pi^{2} m^{2}\omega}
\int_{\Delta\epsilon=\omega}\frac{dS}{|\nabla_\textbf{k}(\epsilon^{+}-\epsilon^{-})|}|\textbf{p}_{+,-}|^{2}
\label{eq:conduct2}
\end{equation}
where ${|\textbf{p}_{+,-}|}$ is the matrix element for transitions between the two bands under consideration.

In the case of a spherical Fermi surface, the integrand is constant and equation (\ref{eq:conduct2}) simplifies to 
\begin{equation}
\sigma_{pam}(\omega) = \frac{e^{2}}{4 \pi^{2} m^{2}\omega}
\sum_{k'=k_< ,k_>}\frac{4 \pi k^{2}|\textbf{p}_{+,-}|^{2}}{|\partial_{k}(\epsilon^{+}-\epsilon^{-})|}\Bigg{|}_{k=k'}
\label{eq:conduct3}
\end{equation}
It is useful to consider the case of constant optical matrix elements ${|\textbf{p}_{+,-}|}$, meaning that every filled \textbf{k} state transits to the corresponding upper band state with equal probability. One can now obtain a model lineshape from Equation \ref{eq:conduct3} using the explicit PAM dispersion relations\cite{grewe84,hewson,millis87a}
\begin{equation}
\label{eqn:epm}
\epsilon^{\pm}=\frac{E_{F}+\tilde{\epsilon_{f}}+\epsilon_{\textbf{k}}\pm \sqrt{(E_{F}+\tilde{\epsilon_{f}}-\epsilon_{\textbf{k}})^{2}+4 \tilde{V}^{2}}  }{2},
\end{equation}
where $E_F$ is the Fermi level, $\tilde{V}$ is the renormalized hybridization strength and ${\tilde{\epsilon_{f}}}$ is the \f-level position renormalized by on-site \f-electron repulsion. This latter quantity defines the scale of the low-energy physics and is commonly identified with the impurity Kondo temperature, \tk\ (discussed further below).

\begin{figure}
\begin{center}
\includegraphics[width=3.4in]{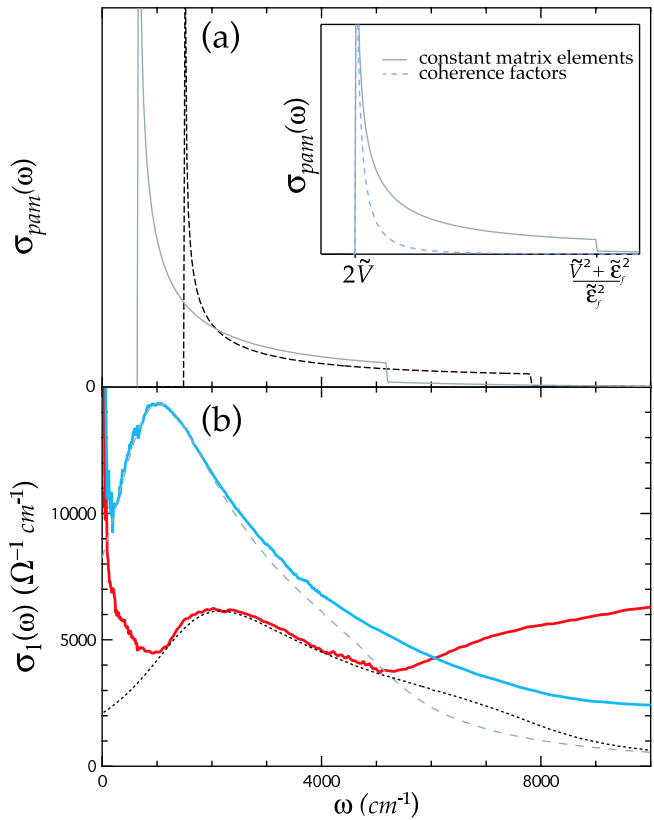}
\caption{(Color online) (a) The idealized conductivity (Equations \ref{eq:conduct4} and \ref{eq:conduct5}) for $\tilde{V}=93\s meV$ and $\tilde{\epsilon_f}=9\s meV$ (solid) and $\tilde{V}=40\s meV$ and $\tilde{\epsilon_f}=2.5\s meV$ (dashed). Inset contrasts the lineshapes derived with constant matrix elements $|\textbf{p}_{+,-}|$ (Equations \ref{eq:conduct4} and \ref{eq:conduct5}), and those with the coherence factors discussed in the text.
%\EF=2\s\eV\ in each case.
(b) The same idealized conductivity curves as (a), Lorenztian broadened with widths $\Delta_0=0.16\s eV$ and $\Delta_{0.75}=0.125\s eV$ for comparison to the measured conductivity of \ybinag\ with \x=0 and \x=0.75, respectively.}
\label{fig:fitplot}
\end{center}
\end{figure}

Putting Equation \ref{eqn:epm} into \ref{eq:conduct3}, the conductivity takes the form:
\begin{equation}
\sigma_{pam}(\omega) = \frac{2 e^2|\textbf{p}_{+,-}|^{2}}{m^2 |\nabla \epsilon_{k_F}|^3\pi}\frac{(E_F+\tilde{\epsilon_f})^2+\omega^2-4\tilde{V}^2}{\sqrt{\omega^2-4\tilde{V}^2}}
\label{eq:conduct4}
\end{equation}
for $2\tilde{V}<\omega<\frac{\tilde{V}^{2}+\tilde{\epsilon_{f}}^{2}}{\tilde{\epsilon_{f}}}$ and 
\begin{equation}
\sigma_{pam}(\omega) = \frac{e^2|\textbf{p}_{+,-}|^{2}}{m^2 |\nabla \epsilon_{k_F}|^3\pi}\frac{(E_F+\tilde{\epsilon_f}-\sqrt{\omega^2-4\tilde{V}^2})^2}{\sqrt{\omega^2-4\tilde{V}^2}}
\label{eq:conduct5}
\end{equation}
for $\omega>\frac{\tilde{V}^{2}+\tilde{\epsilon_{f}}^{2}}{\tilde{\epsilon_{f}}}$. Figure \ref{fig:fitplot}a shows this lineshape for two sets of $\tilde{V}$ and $\tilde{\epsilon_{f}}$ values.

The rather idealized lineshape generated by these considerations is extremely sharp and a meaningful comparison with the data requires addressing the effects of relaxation, which were thus far neglected in our treatment. To this end, we convolute this idealized lineshape with a Lorentzian function, keeping the half width $\Delta$ as an adjustable parameter when performing fits to the measured conductivity. Examples of fits produced using this procedure are shown in Figure \ref{fig:fitplot}b.

The broadening parameter $\Delta$ addresses the finite width of the spectral function along the dispersion curves of Figure \ref{fig:qpd}, and therefore provides a measure of the statistical time over which a typical quasiparticle decays. The numerical values for $\Delta$ obtained from our fits range from 0.12\s\eV\ to 0.17\s\eV, with an associated time scale for decay in the range $\tau=\hbar/\Delta$=5.4\s$ps$ to 3.8\s$ps$, respectively. These lifetimes estimates are in good agreement with the quasiparticle lifetime of members of  this class of materials (\ybag), as measured directly by Demsar \etal\ in pump-probe experiments\cite{demsar} of electron-hole relaxation lifetime. The extraction of this parameter from the conductivity data is a meaningful consistency check on the method developed here.

We now take a moment to consider the possible influence of \textbf{k}-dependent matrix elements associated with the composite nature of the hybridized quasiparticles on the electrodynamic response.
The hybridizing quasiparticles are composite admixtures of excitations with both \f\ and conduction electron character. The regions of the dispersion which are flatter correspond to quasiparticles with a large amplitude of \f\ admixture whereas regions which follow more closely the bare conduction dispersion are dominated by conduction character. Generally speaking transitions among the bare states are not all equally probable (\textit{i.e.} $|\textbf{p}_{cc}|\neq|\textbf{p}_{cf}|\neq|\textbf{p}_{ff}|$). Thus one expects that the optical transition rate for quasiparticles may exhibit some dependence on \textbf{k}, which goes beyond our earlier assumptions.

One can obtain a relatively simple model with \textbf{k} dependence by assuming transitions among bare states only occur between conduction electron initial and final states ($|\textbf{p}_{cc}|\neq0, |\textbf{p}_{cf}|=|\textbf{p}_{ff}|=0$), and using that to calculate transition rates between the hybridized quasiparticle bands. To model the non-constant admixture of states, we use the coherence factors of the resonant level model\cite{tannous,coxcomm} (aka Fano-Anderson\cite{mahan}),
\begin{eqnarray}
\label{eqn:uv}
u_{\textbf{k},\sigma}=\frac{1}{\sqrt{1+(\frac{\tilde{V}}{\tilde{\epsilon_f}-\epsilon^+})^2}},&
v_{\textbf{k},\sigma}=\frac{1}{\sqrt{1+(\frac{\tilde{V}}{\tilde{\epsilon_f}-\epsilon^-})^2}}.
\end{eqnarray}
The approach follows as before however the $JDOS$ integral (Eqn. \ref{eq:conduct2}) picks up a factor $u_{\textbf{k},\sigma}^2v_{\textbf{k},\sigma}^2=\tilde{V}^2/\omega^2$ associated with these coherence factors. One thus obtains a model conductivity similar to Equations \ref{eq:conduct4} and \ref{eq:conduct5}, but with $|\textbf{p}_{+,-}|^2$ replaced by $|\textbf{p}_{cc}|^2\tilde{V}^2/\omega^2$. The most significant effect of the inclusion of coherence factors is that the conductivity at high frequency should fall to zero much more quickly than in the constant matrix element case\cite{coxcomm}. The line shapes with and without coherence factors are contrasted in the inset of Figure \ref{fig:fitplot}. 

This approach may be too nuanced because in a general mixed valent system, bare \f-to-conduction state transitions can be appreciable, and in fact are expected to be important in \ybinag. This is because the conduction band states are derived primarily from Cu-In-Ag $p$ and $d$ orbitals\cite{antonov}, whereas the \f\ electrons sit on Yb sites. This physical displacement between the underlying orbital states is manifest in the banded states through nonvanishing dipole transition matrix elements, $|\textbf{p}_{fc}|$\cite{coxcomm}. Thus, in \ybinag, transitions involving the flatter portions of the quasiparticle dispersion, which are dominated by \f-like character, \textit{can} be expected to provide a considerable contribution to the optical strength, thus we feel that Equations \ref{eq:conduct4} and \ref{eq:conduct5} are more applicable to \ybinag.

\section{Discussion, \tk\ Scaling}
\label{sec:fs}
\begin{figure}
\centerline{\scalebox{.5}{\includegraphics{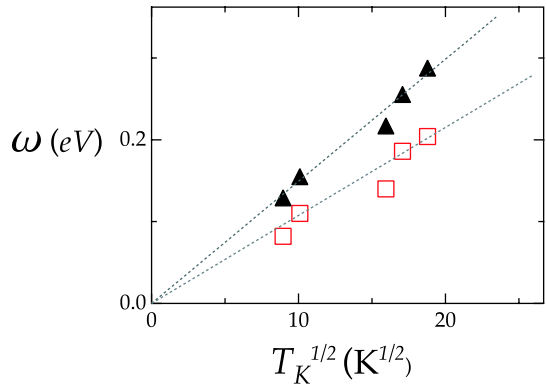}}}
\caption{(Color online) Plot of ${\omega_{pk}}$ and ${\omega_{th}}$ versus ${\sqrt{T_{K}}}$. The dotted lines represent Eq. (\ref{eq:tkb}) and the slopes are ${1.6\hspace{.05cm}eV^{\frac{1}{2}}}$ and  ${1.1\hspace{.05cm}eV^{\frac{1}{2}}}$.} \label{fig:freqscale}
\end{figure}

We now analyze the \x\ dependent frequency of the 2000\s\wn\ feature (Figure \ref{fig:ntw}a), and its relationship to the Kondo temperature \tk\ (Figure \ref{fig:ntw}b). Figure \ref{fig:freqscale} shows $\omega_{pk}$, the maximum of the conductivity, and the threshold frequency $\omega_{th}$, determined from the fit described above, plotted versus the square root of the Kondo temperature. The complex \x\ dependence of $\omega_{th}$ and $\omega_{pk}$ (Figure \ref{fig:ntw}) simplifies considerably when we plot these quantities as a function of \tk\ (Figure \ref{fig:freqscale}). The emergence of a functional relationship between these quantities implies that the frequency of the 2000\s\wn\ peak is controlled by the same physics that underlies the thermodynamic behavior. The square root dependence is evidence that hybridization physics plays a dominant role.

With that in mind, the modeling developed in the previous section can be used to extract an estimate for the band parameter $B$ (Equation \ref{eq:tkb}) by associating the threshold frequency, $\omega_{th}$, with its PAM value, $2\tilde{V}$. The slope of the line through the $\omega_{th}$ values in Figure \ref{fig:freqscale}, together with Equation \ref{eq:tkb}, directly gives $B=0.30\s eV$. This value of $B$ reflects the rate at which the frequency increases with \tk. This can be compared with expectations based on density of states, as well as the rate at which the strength decreases with \tk, as discussed below.

\begin{figure}
\begin{center}
\includegraphics[width=3.2in]{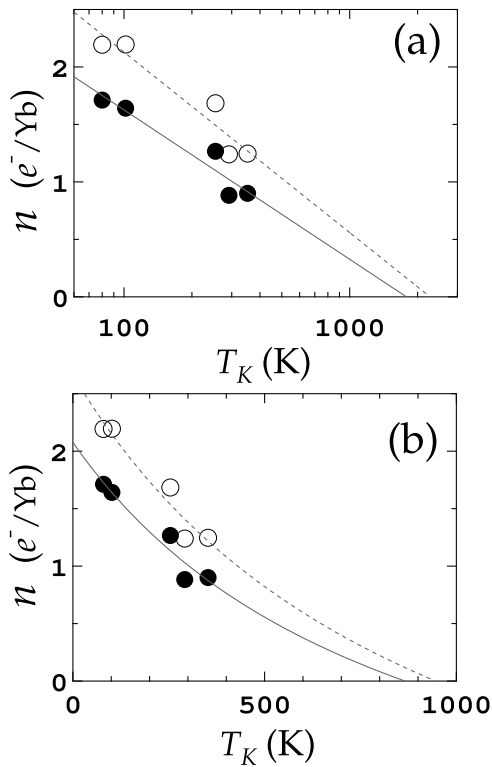}
\caption{Scaling relations for the strength of the 2000\s\wn\ feature with \tk. Dark circles represent $n(4000\s cm^{-1})$ and open circles represent $n(6000\s cm^{-1})$
(a) shows the result of fitting the measured dependence with Equation \ref{eqn:ntknocf2}, and (b) shows the same data fit with Equation \ref{eqn:ntkcf}.}
\label{fig:ntk}
\end{center}
\end{figure}

Within the approach developed in Section \ref{sec:kg} the relationship between \tk\ and the strength of the 2000\s\wn\ feature ($n$) can be addressed. We can obtain a closed-form result from our model calculation if we set the two sphere areas in (\ref{eq:conduct3}) equal to ${4\pi k_{F}^{2}}$. This approximation\footnote{In regards to the \tk\ dependence, this approximation is equivalent to assuming that \EF\ is the largest energy scale in the problem.} avoids the effects of bare band structure details while including the influence of the strong cusp at ${2\tilde{V}}$, which originates from the many-body physics of the PAM. Solving for the wavevectors ${k_{>}}$ and ${k_{<}}$ using the condition ${\epsilon^{+}-\epsilon^{-}=\omega}$, and substituting the result into \ref{eq:conduct3}, we obtain a model strength:
\begin{eqnarray}
n_{pam} & = & \frac{2m}{\pi e^2}\int_{2\tilde{V}}^{\Omega_{FS}}\sigma_{pam}(\omega) d\omega   \\
& \simeq & \frac{4|\textbf{p}_{+,-}|^{2} k_{F}}{\pi^2}\ln\Big(\frac{\tilde{V}}{\tilde{\epsilon_{f}}}\Big)
\label{eqn:ntknocf}.
\end{eqnarray}
This is essentially the area under the curves of Figure \ref{fig:fitplot}a. Using Equation \ref{eq:tkb} and introducing the parameter $c$, defined by $T_K=c\s \tilde{\epsilon_f}$, we can express $n_{pam}$ in terms of \tk:
\begin{eqnarray}
n_{pam} & \simeq & \frac{4|\textbf{p}_{+,-}|^{2} k_{F}}{\pi^2}\ln\Big(c\sqrt{\frac{B}{T_{K}}}\Big)\label{eqn:ntknocf2}.
\end{eqnarray}
Figure \ref{fig:ntk}a shows a least squares fit of this logarithmic scaling relationship to the data. 

The factor $c$ relates the renormalized \f\ level position at low energies and the Kondo temperature, \tk. The value of $c$ is unambiguous in the Fermi liquid theory of the $N(=2j+1)$-fold degenerate Anderson impurity model where,
\begin{equation}
\label{ }
\tilde{\epsilon_f}=\frac{T_K}{c}=k_BT_L\frac{N^2\sin(\pi/N)\cos(\pi/N)}{\pi(N-1)}
\end{equation}
and\footnote{$T_L$ here is a common alternative definition of the Kondo temperature\cite{rajan,cornelius97} and is related to the Kondo temperature $T_K$ by\cite{hewson} Equation \ref{eqn:tktl}.}
\begin{equation}
\label{eqn:tktl}
T_K=w_NT_L.
\end{equation}
$w_N$ is the generalized Wilson number\cite{hewson} given by
\begin{equation}
\label{eqn:wn}
w_N=\frac{e^{1+C-\frac{3}{2N}}}{2\pi\Gamma(1+\frac{1}{N})}.
\end{equation}
For the $j=\frac{7}{2}$ moment of Yb, $N=8$ and $c\simeq0.66$. 

Using this value of $c\simeq0.66$ and fitting the measured result to Equation \ref{eqn:ntknocf} gives $B=0.35\s eV$ and 0.45\s\eV\ for $n(4000\s cm^{-1})$ and $n(6000\s cm^{-1})$ (Eq. 1), respectively. We consider these to be a reasonable agreement given the simplicity of our approach. These values of $B$, which are determined by the amount that $n$ goes up when \tk\ goes down, are similar in size to our previous estimate of the parameter $B=0.30\s eV$ determined from the amount that $\omega_{th}$ goes down as \tk\ goes down. Therefore, in addition to the agreement of experiment and theory regarding the \textit{direction} of the \tk\ dependence of $n$ and $\omega_{th}$, the \textit{sensitivity} of the dependence of $n$ and $\omega_{th}$ on \tk\ are in reasonable agreement \textit{with each other}. In addition, the numerical values for $B$ are reasonable bandwidths for the InAgCu $d$-orbital derived band states\cite{ibach,antonov} of \ybinag. Thus we conclude that the observed dependences of both $n$ and $\omega_{th}$ on \tk\ is consistent with the predictions of the periodic Anderson model in magnitude as well as direction.

An alternative strength estimate can be made using the coherence factor model of the conductivity, introduced in detail above. In that case, we replace $|\textbf{p}_{+,-}|^2$ by $|\textbf{p}_{cc}|^2\tilde{V}^2/\omega^2$ in Equation \ref{eq:conduct2} and repeat the steps which produced Equation \ref{eqn:ntknocf2}, giving a theoretical strength
\begin{eqnarray}
n_{pam} & = & \frac{|\textbf{p}_{cc}|^{2} k_{F}}{\pi^2}\frac{c^2B-T_K}{c^2B+T_K}.
\label{eqn:ntkcf}
\end{eqnarray}
Again using $c=0.66$, fits to the data would yield $B$ estimates $0.17\s eV$ and 0.19\s\eV, for $n(4000\s cm^{-1})$ and $n(6000\s cm^{-1})$, respectively. These values are in order-of-magnitude agreement with the band parameter estimates made above. This functional form is fit for comparison to the data in Figure \ref{fig:ntk}b. However, as discussed at the end of section \ref{sec:kg}, the \f-to-conduction electron transitions are made allowed in this system and so the calculation which includes these, represented by Equation \ref{eqn:ntknocf} is probably more appropriate for \ybinag.

\section{Discussion, High Frequency Interband Transitions}

\begin{figure}
\begin{center}
\includegraphics[width=3.4in]{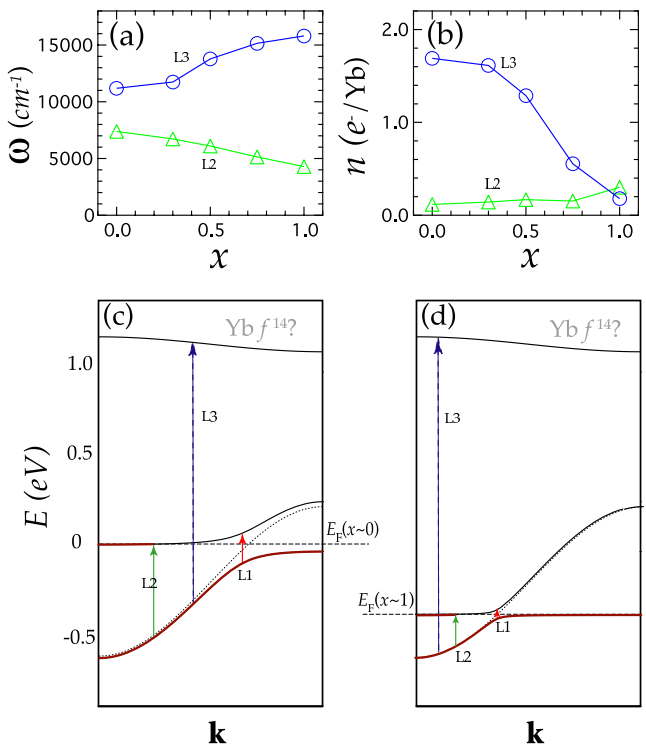}
\caption{(Color online) (a) The center frequencies of the transitions L2 and L3 from (Figure 5), and (b) the strengths of the corresponding Lorentzian fit components are shown as a function of doping, \x. A band-structure picture which we use to interpret the trends in (a) and (b) is shown for (c) \ybin\ and (d) \ybag. The dashed lines illustrate \textbf{k}-conserving transitions which we associate with L1, L2 and L3,
respectively. (L1 is the Kondo resonance excitation.)}
\label{fig:ib}
\end{center}
\end{figure}

%\begin{figure}\begin{center}\includegraphics[width=3.in]{centfreqwps.eps}\caption{(a) The center frequencies and (b) the integrated strengths of the Lorentzian fit components L2 and L3 versus \x.}\label{fig:cf}\end{center}\end{figure}

We now consider the \x-dependent trends in the high frequency ($\omega>6000\s cm^{-1}$) conductivity. We will use the language introduced in Figure \ref{fig:allx} (Section \ref{sec:lo}) associated with Lorentzian fitting of the conductivity data, focussing our attention on the features L2 and L3.  In Figures \ref{fig:ib}a and \ref{fig:ib}b we show the center frequency and strength of these high energy excitations as a function of \x . This measured \x-dependence can inform
our understanding of the nature of the underlying states associated with these transitions and the density of
states near \EF . 

In a textbook picture of metals and semiconductors, an important effect of doping is to add or remove electrons from a set of band states, thereby influencing the position of \EF. In \ybinag, increasing \x\ from 0 to 1 corresponds to the net removal of 2 electrons (per formula unit) from the system, hence we expect the Fermi level to move downward in energy as \x\ is increased. Relativistic band structure calculations addressing these changes have been carried out by Antonov \etal\cite{antonov}. In this scenario, optical features involving transitions from filled states just below \EF\ tend to weaken and move upward as \x\ is increased and these states are emptied\footnote{Similarly, optical features involving final states just above \EF\ may strengthen as \x\ is increased and new final states become available.}. The dashed arrow of Figure \ref{fig:ib}c illustrates one such transition, which we identify with L3. In this scenario, the width of the feature when \x=0 suggests bandwidths of order \twiddle1\s\eV, and the threshold for these transitions, approximately 7000\s\wn\ (\twiddle 0.9\s\eV) when \x=0, gives an indication of the overall energy position relative to \EF. Furthermore, the amount of shift with doping implies that the density of states in the near-\EF\ region of the band structure is approximately 2/0.58\s\eV\twiddle3.5$e^{-}eV^{-1}$/f.u. Both of these numbers are quite reasonable for conduction bands in rare earth and transition metal systems\cite{ibach}.

With this rigid band interpretation as a backdrop describing the salient relationships between L3 and \x, the correlated electron effects discussed in previous sections occur in addition. The renormalization discussed there dresses these bare states and the renormalized region of the band structure, including the Kondo resonance, tracks \EF, which moves downward with increasing \x.

We now turn to the systematics of the feature L2. This feature redshifts with \x\ by an amount similar in magnitude to the shift of L3 (Figure \ref{fig:ib}a). However, the narrowness and nearly \x-independent strength of L2 does not lend as easily to a simple band interpretation. When considering an identification of the component L2, we point out the significant temperature dependence in the frequency region associated with L2, as shown before in Figures \ref{fig:x0}, \ref{fig:x3} and \ref{fig:x1}. The temperature dependent interplay of spectral weight contained in the frequency intervals of L1 and L2 naturally lead one to speculate that perhaps the same correlated electron physics controlling L1 may also be relevant to L2. One is thus led toward the question of whether the presence of the feature L2 represents a further phenomenon associated with hybridization physics. 

We noted in Section \ref{sec:pam} that in addition to the strong peak in the PAM conductivity associated with renormalized band nesting, a second feature could appear at higher frequencies associated with the initial state energy crossing \EF. In a linearly dispersing band model, this change occurs around the conduction electron bandwidth frequency (Equation \ref{eqn:OFS}). In a more realistic bandstructure, the filling fraction and band curvature details could influence the frequency and magnitude of this conductivity change. In particular, if the transition final states are band edge states, then significant shifting with doping concentration could result.

With these considerations, it is reasonable to suggest that L2 represents a conductivity feature arising from Fermi surface quasiparticles. The narrowness can then be attributed to the very long lifetime of the initial state (because it occurs on the Fermi surface), and weak \x\ dependence of the strength arises because the energy structures responsible for the associated transitions track the \x\ dependent Fermi level, as opposed to becoming filled or depleted with \x. The culmination of the identifications suggested in this section are presented at a light and heavy doping concentration in Figures \ref{fig:ib}c and \ref{fig:ib}d, respectively.

\section{Discussion, The Phase transition of Y\lowercase{b}I\lowercase{n}C\lowercase{u}$_4$}

The physical mechanism of the phase transition in the lightly doped system remains elusive, however, progress has been made in understanding aspects of the phase transition.
In previous sections, we have analyzed in detail the low temperature conductivity in terms of PAM renormalizations of few-band models. We now discuss the how this picture identifying optical transitions may be related to the interesting phenomenology displayed by \ybinag\ including the phase transition at the low \x.

The Kondo volume collapse\cite{allen1} (KVC) model, which describes the complex interplay of the system volume, hybridization, Kondo temperature, and \f\ level occupation, seems adequate to describe the valence transition in elemental Ce\cite{allen1,haule}. This model, however, seems insufficient to quantitatively describe the phase transition in \ybin, as evidenced mainly by the smallness of the volume change at the transition. Some authors\cite{sarrao3,cornelius97,figueroa98} have argued that a quasigap, or region of low density of states, exists in the bare band structure located just above the Fermi level in \ybin, and is important to the phase transition. This description is qualitative, but addresses the change in carrier density as well as the changes related to Kondo physics.
In this approach, the presence of a quasigap makes the Kondo temperature very sensitive to the placement of the Fermi level and can help induce the phase transition in manner akin to the KVC model, but with the importance placed on the density of states dependence, rather than the volume.

It is of interest to ask what consequences a quasigap scenario could have for the optical data in \ybin, where the conductivity in the range of 8000\s\wn\ (L2 in the fits) drastically displaces to lower frequency forming the Kondo resonance excitation (L1) discussed above. In the picture outlined in Figure \ref{fig:ib}c, an upward shift in the Fermi level off of the bare band edge and into the quasigap could have drastic consequences for the optical response, possibly forcing the contribution L2 to shift to very high frequency, as the renormalized portion of the upper band is forced across the gap in response to a small shift in \EF. This possible interpretation of the temperature dependence indicates that further theoretical work directed toward investigating the generic physics of the PAM and the fate of the Kondo scenario in the context of rapidly varying band structure is needed to understand the complex temperature dependence of \ybin.

Theoretical work on the Falicov-Kimball model\cite{zlatic01,freericks03,fkrmp} attempts to make a quantitative connection with experiment using a particular many-body model which features a first-order phase transition. Aspects of this modeling seem promising, in particular the prediction of substantial temperature dependence of the high frequency optical conductivity, however this model does not yet explicitly include the effects of hybridization. It is possible that a minimal model that describes the first order phase transition and also includes the Kondo physics may require and extension of the periodic Anderson model to include Falicov-Kimball-type interaction.

\section{Conclusions}

Our results indicate that \tk\ scaling is present in the low temperature finite-frequency dynamics of \ybinag\ and can be addressed in the context of local-moment models. Furthermore, our data provide numerical estimates of key parameters necessary for the construction of a minimal theoretical model of the valence transition as well as pointing out salient features of the underlying bandstructure. Further work may be directed toward greater understanding of this low-$T$ scaling behavior as well as unexplained temperature dependent behavior of lightly doped \ybinag.

\begin{acknowledgments}
The authors have greatly benefitted from discussions with D. N. Basov, A. L. Cornelius, D. L. Cox, P. A. Lee, B. S. Shastry, and A. P. Young. We also gratefully acknowledge S. L. Hoobler and Y. W. Rodriguez for technical assistance. Work at UCSC supported by NSF Grant Number DMR-0071949. ZF acknowledges support of NSF Grant Number DMR-0203214.
\end{acknowledgments}

\bibliography{zs-short,ybcu,ybcu04,valence,kondo03,jasonbooks}

\end{document}